\documentstyle[prb,aps,12pt]{revtex}

\begin{document}
\title{Coexistence of large and small polarons in manganites}
\author{V.~Cataudella, G.~De~Filippis and G.~Iadonisi}
\address{INFM, Unit\`a di Napoli, Dipartimento di Scienze Fisiche, \\
Universit\`a di Napoli I-80125 Napoli, Italy}
\date{\today}
\maketitle

\begin{abstract}
The interplay of the electron-phonon interaction and of the double and super
exchange magnetic effects is analyzed in the $La_{1-x}Ca_xMnO_3$ perovskites
with $0<x<0.5$. By using an analytical variational scheme that allows to
treat the electron-phonon interaction in a fully quantum manner it is shown
that in the intermediate electron-phonon coupling regime a charge density
instability occurs near the metal-insulator transition, induced by either
temperature or hole doping. The system segregates in antiferromagnetic or
paramagnetic small polarons and ferromagnetic large polarons domains
characterized by different values of the lattice distortions that are
satisfactory compared with those experimentally observed.
\end{abstract}


\newpage {\bf Introduction}. In recent years the perovskite oxides $%
La_{1-x}Ca_xMnO_3$ have become the focus of the scientific interest owing to
the discovery of the colossal magneto resistance effect. These materials
have been first studied in the 1950's for their peculiar phase diagram where
is present a strong correlation between magnetization and resistivity.\cite
{1} In particular, in the phase of colossal magneto resistance,\cite{1bis}
characterized by $0.2<x<0.5$, these compounds exhibit a transition from a
paramagnetic insulator to a ferromagnetic metal upon cooling that has been
generally explained by the double exchange mechanism originally introduced
by Zener.\cite{2} The discovery of the colossal magneto resistance
phenomenon and the observation, by means of the resistivity studies,\cite{3}
that the tight binding band description of the conduction electrons\cite{2,4}
is inadequate and that strong interactions between the electrons and the
lattice distortions play a non negligible role have aroused a renewed
interest for these compounds. The relevance of the Jahn-Teller polaron
formation, first suggested by Millis,\cite{3} has been shown by giant
isotope shift of the Curie temperature,\cite{5} by measurements of the
lattice distortions by EXAFS\cite{6} and by frequency shifts of the internal
phonon modes\cite{7} and now it is generally recognized.\cite{8}

Recently, a large amount of experimental data has been collected near the
phase transition boundaries induced by either temperature or hole doping.%
\cite{9}  The analysis of the experimental data showed the presence of 
mixed phases.  For compositions intermediate near to the metal-insulator
transition at low temperatures the ferromagnetic and antiferromagnetic
phases coexist while  the ferromagnetic and paramagnetic ones coexist in the
vicinity of the paramagnetic to ferromagnetic transition temperature. On the
other hands the existence of a homogeneous canted phase first proposed by de
Gennes\cite{10} in the attempt of explaining the weak ferromagnetism of the
phase with $0<x<0.2$ has not been confirmed. 

In these materials, the ferromagnetic phase is induced by the combined
effect of the $e_g$ electron hopping between nearest neighbor sites and
Hund' s exchange with the localized $t_{2g}$ electron spins: this phase is
associated with the properties of a metallic state. On the contrary the
antiferromagnetism is induced by the exchange integral between the $d$
shells of adjacent atoms and it is connected with the properties of an
insulator state.

In this paper we study, on the basis of a variational approach, the
interplay of the electron-phonon interaction and of the double and super
exchange magnetic effects as function of the doping and the temperature in
the range $0<x<0.5$. We will show that in the intermediate electron-phonon
coupling regime the combined effect of the magnetic and electron-phonon
interactions pushes the system towards a charge density instability. This
explains the experimentally observed tendency of the manganites to form
inhomogeneous magnetic structures near the phase boundaries due to
temperature or $Ca$ concentrations. The metal-insulator transition in these
systems is a first order transition\cite{goode} accompanied by the
coexistence of two phases: a high density one made by localized small
polarons forming paramegnetic or antiferromagnetic domains depending on
temperature and a low density one made by itinerant large polarons forming
ferromagnetic domains.

From a theoretical point of view the coexistence between hole-undoped and
hole-rich phases has been discussed by using exact numerical approaches on
small lattices treating the Jhan-Teller phonons classically\cite{moreo}. 
However, little attention has been given to the quantum nature of
the Jhan-Teller phonons, and, more important, to the possible formation of
polarons characterizing the lattice deformations associated to the phase
separation. On the other hand, after Millis et al.\cite{4}, the role of
polarons in the manganites has been clearly recognized. The main goal of the
present letter is to show that the two scenario are fully compatible and
that the formation of two types of polarons is responsible, together with
the magnetic interactions, of the complexity of the manganites in the
region $0<x<0.5$.

{\bf The model}. In $La_{1-x}Ca_{x}MnO_3$ the electronically active orbitals
are the Mn $3d$ orbitals and the mean number of the $d$ electrons per $Mn$
is $4-x$. Three electrons go into the $t_{2g}$ core states and the remaining 
$1-x$ electrons occupy the outer shell $e_g$ orbitals. The Jahn-Teller
effect splits the $e_g$ double degeneracy\cite{13} and the single orbital
approximation is reasonable for $x<0.5$. The corresponding Hamiltonian is
given by: 
\begin{eqnarray}
H=&&-t\sum_{i,j} \left(\frac{S^{i,j}_0+1/2}{2 S+1}\right)
c^{\dagger}_{i}c_{j} +\omega_0\sum_{\vec{q}}a^{\dagger}_{\vec{q}}a_{\vec{q}}
+\epsilon\sum_{i,j} \vec{S}_{i} \cdot \vec{S}_{j}  \nonumber \\
&&+\frac{g \omega_0}{\sqrt{N}} \sum_{i,\vec{q}} c^{\dagger}_{i}c_{i}\left(a_{%
\vec{q}} e^{i \vec{q} \cdot R_i} +a^{\dagger}_{\vec{q}} e^{-i \vec{q} \cdot
R_i}\right).  \label{1r}
\end{eqnarray}

Here the first term describes the double exchange mechanism, in the limit
where the intra-atomic exchange integral $J$ is much greater than the
transfer integral $t$, $S^{i,j}_0$ represents the total spin of the
subsystem consisting of two localized spins on sites $i$ and $j$ (nearest
neighbors) and the conduction electron, $S$ is the spin of the $t_{2g}$ core
states, $\omega_0$ denotes the frequency of the optical local phonon mode,
the dimensionless parameter $g$ indicates the strength of the
electron-phonon interaction in the Holstein model,\cite{12} $\epsilon$
represents the antiferromagnetic super-exchange coupling between two nearest
neighbor $t_{2g}$ spins and $N$ is the number of lattice sites. The units
are such that $\hbar=1$.

{\bf The variational approach}.

We perform two successive canonical transformations. The first is a
generalization of the Lang-Firsov unitary transformation\cite{14}: 
\begin{equation}
S_{1}=exp\left[-\frac{1}{\sqrt{N}}\sum_{j,\vec{q}} c^{\dagger}_{j}c_{j} e^{i%
\vec{q}\cdot \vec{R}_{j}} f_{\vec{q}}\left(a_{\vec{q}}-a^{\dagger}_{-\vec{q}%
}\right)\right]  \label{2r}
\end{equation}
where the phonon distribution function $f_{\vec{q}}$ has to be determined
variationally.

This choice is related to the approximation first used by Tomonaga\cite{15}
in the treatment of the coupling between mesons and nucleons and based on
the assumption that all mesons are virtually emitted independently of the
others. It recovers the well known Lang-Firsov\cite{14} canonical
transformation assuming narrow band small polaron, and, moreover, it
provides, in the weak coupling regime, the generalization to the many-body
problem of the unitary transformation utilized by Lee-Low-Pines\cite{16} in
the continuum approximation within the Frohlich model.\cite{17}

The second canonical transformation is a Bogoliubov transformation. It
introduces correlations between the emission of successive virtual phonons
by the electron and it is responsible of the phonon frequency renormalization%
\cite{18}: 
\begin{equation}
S_{2}=exp\left[-\alpha\sum_{\vec{q}}\left( a^{\dagger}_{\vec{q}%
}a^{\dagger}_{-\vec{q}} -a_{-\vec{q}}a_{\vec{q}}\right)\right]  \label{3r}
\end{equation}
where $\alpha$ indicates a variational parameter.

The above introduced canonical transformations used to treat variationally
the electron-phonon interaction are known to give a satisfactory description
of the single polaron problem in all range of couplings being able to
interpolate from weak (large polaron) to strong (small polaron)
interactions. For this reason we adopt the same approach here for a system
of interacting polarons. However, it is worth to note that, in the present
case, the utilized approach is expected to be even better than in the single
polaron problem. In fact, the choice of richer trial wave function made as a
linear superposition of Bloch states translationally invariant that, in the
single polaron problem, gives a highly accurate description of the polaron
features in the intermediate electron-phonon coupling regime,\cite{us} does
not provide any improvement in a many-polaron system since the asymptotic
wave functions are orthogonal and the off-diagonal matrix elements of the
Hamiltonian are zero in the thermodynamic limit.

The transformed Hamiltonian is: 
\begin{eqnarray}
\bar{H}=&&-t\sum_{i,j} \left(\frac{S^{i,j}_0+1/2}{2 S+1}\right)
c^{\dagger}_{i}c_{j}X^{\dagger}_i X_j +\omega_0 \cosh{4 \alpha} \sum_{\vec{q}%
}a^{\dagger}_{\vec{q}}a_{\vec{q}}  \nonumber \\
&&+\frac{\omega_0}{\sqrt{N}}\sum_{\vec{q},i}c^{\dagger}_{i}c_{i} \left(a_{%
\vec{q}} +a^{\dagger}_{-\vec{q}}\right) \left(g-f_{\vec{q}}\right)e^{i\vec{q}%
\cdot \vec{R}_{i}} e^{2 \alpha}+\omega_0 \sinh{2 \alpha} \cosh{2 \alpha}
\sum_{\vec{q}} \left(a^{\dagger}_{\vec{q}}a^{\dagger}_{-\vec{q}} +a_{-\vec{q}%
}a_{\vec{q}}\right)  \nonumber \\
&&+\omega_0 N \left(\sinh{\ 2 \alpha}\right)^2 +\frac{\omega_0}{N}\sum_{i,j,%
\vec{q}} \left(f^2_{\vec{q}}-2gf_{\vec{q}}\right) \cos{\left[\vec{q}%
\cdot\left(\vec{R}_i-\vec{R}_j\right)\right]} n_{i}n_{j} +\epsilon\sum_{i,j} 
\vec{S}_{i} \cdot \vec{S}_{j}~.  \label{5r}
\end{eqnarray}
where 
\begin{equation}
X^{\dagger}_i X_j=exp\left[\frac{1}{\sqrt{N}}\sum_{\vec{q}} \left(e^{i\vec{q}%
\cdot\vec{R}_j}- e^{i\vec{q}\cdot\vec{R}_i}\right)f_{\vec{q}}e^{-2 \alpha}
\left(a_{\vec{q}}-a^{\dagger}_{-\vec{q}}\right)\right]~.  \label{7r}
\end{equation}
By assuming a simple square of density of states, in the mean field
approximation, the free energy per site becomes: 
\begin{eqnarray}
\frac{F}{N}=&& \pm \frac{\epsilon}{2} Z S^2 m^2_S +K_{B}T \log{%
\left(1-e^{-\beta\omega_0\cosh{4 \alpha}}\right)} +\rho\left(1-\rho\right)
\left[\frac{\omega_0}{N}\sum_{\vec{q}}\left(f^2_{\vec{q}}-2gf_{\vec{q}}
\right)-W\right]  \nonumber \\
&& +\omega_0 \left[\left(\sinh{2 \alpha}\right)^2-\rho^2 g^2\right]
+K_{B}T\left[ \rho \log{\rho}+\left(1-\rho\right)\log{\left(1-\rho\right)}
+\lambda m_S-\log{\nu_S} \right]  \label{8r}
\end{eqnarray}
where the top and bottom sign, in the first term, hold, respectively, for
the ferromagnetic and antiferromagnetic solutions for the localized spins.
In the Eq.(\ref{8r}) $Z$ indicates the number of nearest neighbors, $\rho$
is the electron concentration, $\lambda$ is a dimensionless variational
parameter that is proportional to the effective magnetic field, $\nu_S$ and $%
m_S$ represent, respectively, the partition function and the magnetization
of the localized spins and $W$ denotes the full band width: 
\begin{equation}
W=Z t \langle \left(\frac{S_0+1/2}{2S+1}\right)\rangle e^{-\frac{1}{N}\sum_{%
\vec{q}} f^2_{\vec{q}} \left[1-\cos{\left(q_x a\right)}\right]
\left(2N_0+1\right) e^{-4 \alpha}}  \label{9r}
\end{equation}
where $a$ is the lattice parameter and $N_0$ is the average phonon number
with renormalized frequency $\omega_0\cosh{4 \alpha}$. The factor $\langle
\left(\frac{S_0+1/2}{2S+1}\right)\rangle$ gives an estimate of the reduction
of the hopping due to the double exchange mechanism and it is evaluated in
the ferromagnetic case by following Kubo and Ohata\cite{19} and in the
antiferromagnetic case by making use of the Clebsch-Gordan coefficients.
Finally the expression of the phonon distribution function $f_{\vec{q}}$ is
obtained by minimizing the free energy per site: 
\begin{equation}
f_{\vec{q}}=\frac{g}{1+\frac{W}{\omega_0}e^{-4\alpha}\left(2N_0+1\right)
\left[1-\cos{\left(q_x a\right)}\right]}~.  \label{10r}
\end{equation}

{\bf The results}. In the case of a single electron coupled to an optical
local phonon mode we have shown\cite{us} that a linear superposition of two
Bloch wave functions describing the small and large polarons gives a very
good estimate of the polaron ground state energy and provides an highly
accurate description of the polaron features in the intermediate
electron-phonon coupling regime and for electron and phonon energy scales
not well separated. In particular, in the intermediate polaron the spectral
weight is equally distributed between the ground and all the excited states
and the polaron wave function contains, in equal weights, the small and large
polaron contributions. It is naturally to ask what happens when there are
many polarons since, now, we can expect a phase separation. To this end we
first perform the minimization of the free energy, subject to the condition
that the total number of particles is fixed, with respect to the two
variational parameters $\alpha$ and $\lambda$ by making use of the
self-consistent equation for the bandwidth $W$. Then, in order to
investigate the existence of coexisting phases, we study the function $%
G(\rho,\mu)=F(\rho)-\mu \rho$, where $F(\rho)$ is the free energy calculated
from the above discussed minimization procedure and $\mu$ is a Lagrangian
multiplier to be adjusted to get the desired total number of particles. The
function $G(\rho,\mu)$, at fixed value of the temperature, is reported in
Fig.1 for different values of the chemical potential $\mu$ in the
intermediate electron-phonon coupling regime $\lambda_p \simeq 1$ ($%
\lambda_p $ represents the ratio between the small polaron binding energy
and the energy gain of an itinerant electron on a rigid lattice). From Fig.1
it is evident that there are values of the chemical potential $\mu$ such
that the function $G$ is stationary for three distinct electron densities.
In particular, there is a value of $\mu$ such that $G$ is the same at two
different values of the electron density: $\rho_1$ and $\rho_2$. They
characterize the coexistence region. We stress that $\rho_1$ and $\rho_2$
depend on the temperature and correspond to homogeneous phases of the system
constituted by small and large polarons. Below a critical temperature the
system segregates (see Fig.2) in antiferromagnetic or paramagnetic and
ferromagnetic domains (with density $\rho_1$ and $\rho_2$) characterized by
different values of the lattice distortions and by distinct behaviors of the
resistivity with the temperature. The domains of large polarons are
characterized by a weak electron mass renormalization and by coherent
motion. On the other hand within the domains of small polarons the
well-known band collapse takes place, the coherent motion is suppressed
rapidly with increasing the temperature and the resistivity has a thermal
activated behavior characteristic of the semiconducting phase.

In the region of coexistence the fractions of volume, $V_1/V$ and $V_2/V$,
filled with density $\rho _1$ and $\rho _2$, are determined by the two
following conditions: $V_1/V+V_2/V=1$ and $V_1/V\rho _1+V_2/V\rho _2=\rho $,
where $\rho =n/V$ and $n$ represents the total number of charge carriers. It
is crucial, in order to confirm our results, to evaluate the lattice
displacement associated to the two kinds of coexisting domains. An estimate
of the average deviation of the molecule on the site $i$ from the
equilibrium position when one electron is on site $i$ is given by the
function: 
\begin{equation}
D=\frac 1{N\rho }\sum_i\langle c_i^{\dagger }c_i\left( \frac{%
a_i+a_i^{\dagger }}{\sqrt{2M\omega _0}}\right) \rangle  \label{24r}
\end{equation}
where $M$ denotes the ionic mass.

Applying the two canonical transformations $S_{1}$ and $S_{2}$ one finds: 
\begin{equation}
D=\frac{2}{\sqrt{2 M \omega_0}} \left[\left(1-\rho\right)\frac{1}{N}\sum_{%
\vec{q}}f_{\vec{q}}+g \rho \right]~.  \label{25r}
\end{equation}
In particular, for narrow band small polarons $f_{\vec{q}}$ is of order of $%
g $ and $D$ assumes the value $2g/ \sqrt{2M \omega_0}$.

By using the following values for the model parameters: $t=2\omega_0$, $%
g=3.5 $, $\epsilon=0.01 t$, $\omega_0=50 meV$, for $M$ the mass of the
oxygen atom and by taking into account that the Jahn-Teller distortion of
the $Mn^{3+}$ ion consists in an axial elongation of the two $MnO$ bonds of
the $MnO_6$ octahedra, at $T=240 K$ we find: $D/2 \simeq 0.18~\AA$ for the
paramagnetic domains of small polarons and $D/2 \simeq 0.12 ~\AA$ for the
ferromagnetic domains of large polarons. These results are in agreement with
the experimental data\cite{6} on $La_{0.75}Ca_{0.25}MnO_3$ compound, that
show evidence, above $170 K$, of small polarons characterized by $MnO$ bond
of order of $\sim0.22~\AA$ and large polarons with $MnO$ bond of order of $%
\sim0.1~\AA$.

In summary, we have studied the combined effect of the magnetic and
electron-phonon interactions in $La_{1-x}Ca_{x}MnO_3$ perovskites by means
of two canonical transformations and the mean field theory. It has been
shown that in the intermediate electron-phonon coupling regime these systems
form inhomogeneous magnetic structures near the metal-insulator transition,
segregating into antiferromagnetic or paramagnetic and ferromagnetic domains
of small and large polarons characterized by different values of the lattice
distortions as experimentally observed.

\section*{Figure captions}

\begin{description}
\item  {F1} The function $G(\rho ,\mu )=F(\rho )-\mu \rho $ is plotted as a
function of the doping for different values of the chemical potential at
fixed temperature: $K_BT=0.4\omega _0$ (for $\omega _0=50meV$ $%
K_BT=0.5\omega _0$ corresponds to a temperature T of order of $300K$).

\item  {F2} The phase diagram corresponding to: $t=2\omega _0$, $g=3.5$, $%
\epsilon =0.01t$. The area between the dotted lines indicates the region
where the localized and delocalized phases coexist.
\end{description}

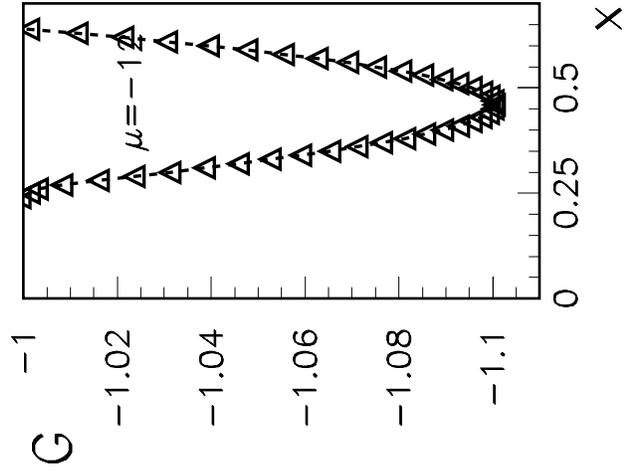
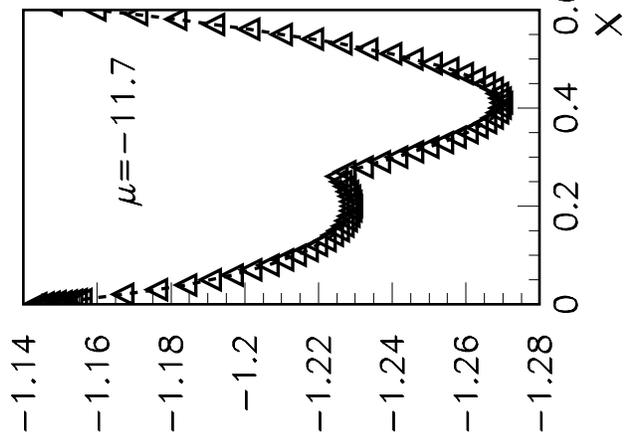
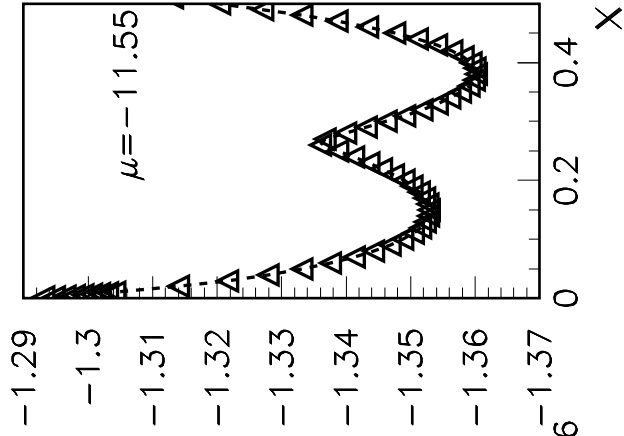
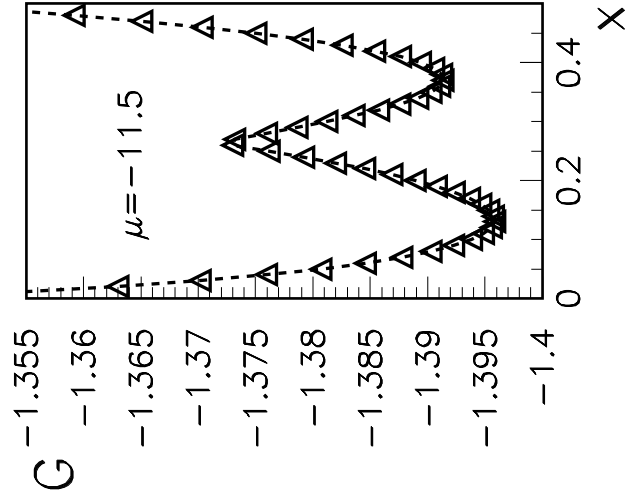
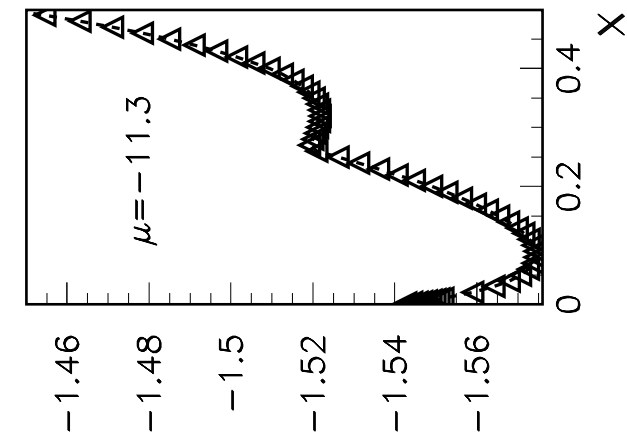
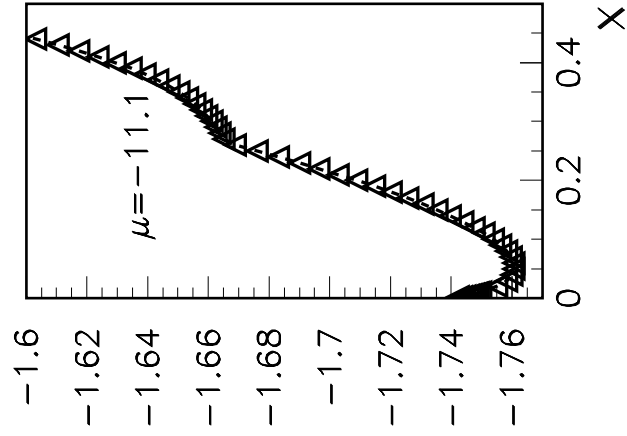
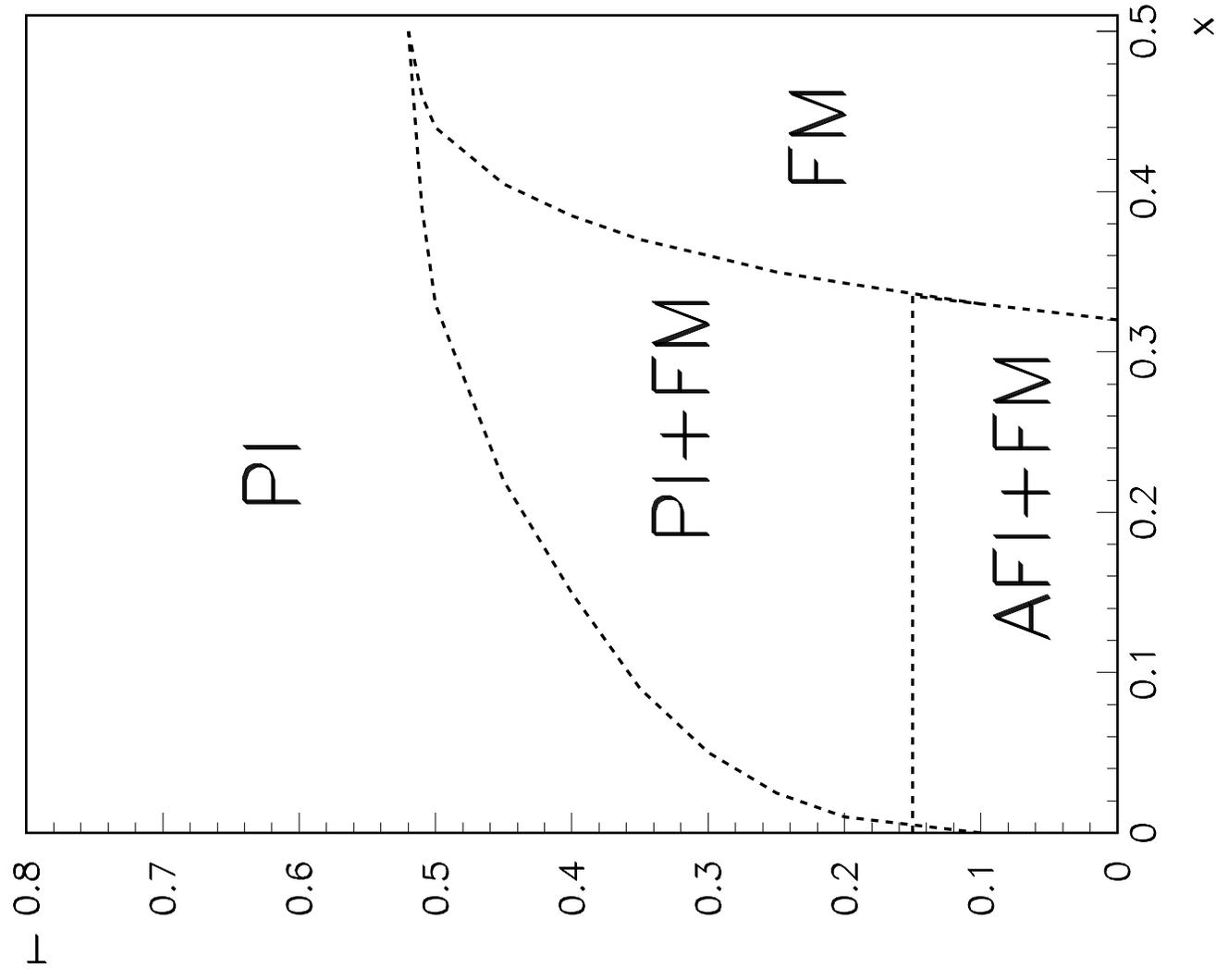
\end{document}